\documentclass[10pt]{iopart} 
\usepackage{graphicx}
\usepackage{color}
\newcommand{\bra}{\langle}
\newcommand{\ket}{\rangle}
 
\newcommand{\threej}[6]{
     \left (\begin{array}{ccc}
              #1 & #2 & #3 \\
              #4 & #5 & #6 
            \end{array}  \right )}

\newcommand{\bff}[1]{\ensuremath{\mathbf{#1}}}

\newcommand{\be}{\begin{equation}}
\newcommand{\ee}{\end{equation}}
\newcommand{\beqa}{\begin{eqnarray}}
\newcommand{\eeqa}{\end{eqnarray}}
\newcommand{\bk}{\mathbf{k}}
\newcommand{\bx}{\mathbf{x}}
\newcommand{\ev}[1] {\left|\left \langle #1  \right \rangle \right|^2}
\begin{document}

\title{Photoassociation of Universal Efimov Trimers}
\author{Betzalel Bazak and Nir Barnea}
\address{The Racah Institute of Physics, The Hebrew University, 
9190401, Jerusalem, Israel}
\date{\today}

\begin{abstract}
In view of recent experiments in ultracold atomic systems, the
photoassociation of Efimov trimers, composed of three identical bosons,
is studied utilizing the multipole expansion. 
We study both the normal hierarchy case,
where one-body current is dominant,
and the strong hierarchy case,
relevant for photoassociation in ultracold atoms, 
where two-body current is dominant.
For identical particles in the normal hierarchy case, 
the leading contribution comes from the 
$r^2$ $s$-mode operator and from the quadrupole $d$-mode operator. 
The $s$-mode reaction is found to be dominant 
at low temperature, while as the temperature increases the $d$-mode becomes 
as significant.
For the strong hierarchy case, the leading contribution comes from a 2-body
$s$-wave $\delta$ operator. 
In both cases log periodic oscillations are found in the 
cross section. For large but finite scattering length the amplitude 
of the oscillations becomes larger in comparison to infinite scattering
length case.
We apply our theory to photoassociation of $^7$Li ultracold atoms and show
a good fit to the available experimental results.
\end{abstract}


\maketitle
\ioptwocol
\section{Introduction}
When the properties of a few-body or many-body system are insensitive to 
the details of the microscopic interaction between its constituents, the system
is said to be {\it Universal}. A free gas is the simplest and somewhat trivial
example of universality. A very rich and important case is a system governed by
the low energy scattering parameters. 
Such scenario can be realized in ultra cold atomic systems, 
where the two-body scattering length can 
be tuned by a Feshbach resonance to be much larger than any other length scale
in the system \cite{BraHam06,ChiGriJul10}.  

The trimer, or the three body system, attracts special attention as 
the simplest non-trivial universal system. 
Moreover, in the 70's Efimov predicted that in the limit of a
resonant 2-body interaction, the system reveals universal properties
\cite{Efi70,Efi71}.
A peculiar prediction is the existence of a series of giant three body
molecules, known as Efimov trimers, that was verified 
experimentally a few years ago. For a review see Ref. \cite {FerZenBer11}
and references therein.

The coupling of a radiation field with the system is a useful experimental tool
in physics. The photoassociation and photodissociation processes are sensitive
ways to measure spectrum and also other properties of few and many-body systems 
\cite{ThoHodWie05}.
Recently, trimers were formed by radio-frequency (rf) excitations 
in both fermionic $^6$Li \cite{LomOttSer10, NakHorMuk11}
and bosonic $^7$Li \cite{MacShoGro12} systems.

In a previous work \cite{BazLivBar12}, 
we have presented the multipole analysis of an rf association process 
binding a molecule of $N$ identical bosons.
We have considered two different scenarios relevant to
photoassociation experiments in neutral ultra cold atoms, confined in a 
strong magnetic field: {\it(i)} spin-flip process, and {\it(ii)} frozen-spin
process. 

In the spin-flip mechanism, the photon induces a spin state change
in the absorbing atom, and most of its energy is dedicated
transferring the atom from one hyperfine state to another. The
typical energy of the photon in this case is of the order of the hyperfine
splitting. The spin-flip mechanism was 
materialized experimentally by \cite{LomOttSer10, NakHorMuk11}.
Previous analysis of rf experiments 
\cite{ChiJul05, BerJesMol06, HanKohBur07, KleHenTop08,TscRit11},
which relied on the Franck-Condon factor, 
is appropriate for describing these spin-flip reactions. 

In the second, frozen-spin, mechanism, the photon energy is insufficient to
bridge the hyperfine gap and therefore the atom spin is ``frozen'' throughout
the process. In this case the photon induces a molecular 
bound-free transition and the typical photon
energy is of the order of the molecular binding energy.

In \cite{BazLivBar12} we have shown that the spin-flip and frozen-spin
processes differ by their operator structure and by the de-excitation
modes that contribute to the photoassociation rate. The photoassociation
experiments carried out by the Bar-Ilan group, see e.g. \cite{MacShoGro12},
relied on the frozen-spin mechanism.

We have studied one-body current in frozen-spin reactions,
dealing with dimer formation \cite{BazLivBar12}, and also calculated 
numerically the quadrupole response of a bound bosonic trimer
\cite{LivBazBar12}. 
Recently we have identified another scenario, where two-body currents dominate
the frozen-spin process,
and explored the occurrence of log-periodic oscillations in the reaction cross 
section \cite{BazBar14}.

Here, we study few more aspects of the three body photoassociation process. 
These are 
{\it (i)} The interplay between the $s$-mode and the $d$-mode reaction channels
at different temperatures and photon energies, 
{\it (ii)} The effect of finite scattering length on the reaction rate.
We utilize the hyperspherical coordinates with the adiabatic expansion to study 
the trimer photoassociation process, for both one-body process
as well as for the two-body process.
Analytic results for the transition rates at the unitary 
point (where the scattering length diverges) 
are derived using the zero-range approximation, and numerical 
calculations complete the picture for a large but finite scattering length.
Similarly to the dimer case \cite{BazLivBar12}, the $s$-mode and the $d$-mode 
are found to be the leading order contributions in the one-body process. 
The relative importance of these modes depends on the temperature, where 
at low temperature the $s$-mode is dominant while at higher temperature 
the $d$-mode becomes as significant.
For both one-body and two-body processes, log-periodic oscillations modulate 
the cross section. We found that comparing with the unitary point,
the amplitude of these oscillations 
is somewhat larger for finite scattering lengths. 

As a concrete example for the usefulness of our approach we analyze 
the dimer and trimer photoassociation experiments carried out 
by the Bar-Ilan group in an ultracold atomic  $^7$Li gas \cite{MacShoGro12}. 
As already pointed out, the photoassociation mechanism in these experiments is
the frozen spin process. 
Due to the strong hierarchy, see Sec. \ref{sec:model},  the coupling of the
photon to the system is dominated by the 2-body magnetization current.
Our theory fits well to the available experimental results.

The paper is organized as follows, in section 2 we introduce our model and 
reiterate the multipole and the currents expansions, describing the interaction 
of an rf field with the atomic system.
In section 3 we introduce the three-body problem and solve it using 
the hyperspherical adiabatic expansion. The transition
matrix elements are calculated in section 4 for infinite,
and also for large but finite, scattering lengths.
The transition rates are calculated in section 5, 
and a detailed comparison to experiment is presented in section 6.
Section 7 concludes our study.

 \section{The model} \label{sec:model}

In photoassociation experiments \cite{LomOttSer10, NakHorMuk11, MacShoGro12}, 
a few MHz rf radiation field is applied to the ultracold atomic gas.
Stimulated emission occurs when the photon energy matches the difference 
between the ultracold gas atoms and the molecule energy, resulting in
molecule formation. 
The transition rate of such process is given by Fermi's golden rule,
\be \label{golden_rule}
r_{i\rightarrow f}=\frac{2\pi}{\hbar} \bar{\sum_i} \sum_f
|\bra f, \bk \zeta | \hat{H}_I | i\ket|^2 
\delta(E_i-E_f-\hbar\omega),
\ee 
where $\bar{\sum_i}$ is an average on the appropriate initial continuum states 
and $\sum_f$ is a sum on the final bound states.
The coupling between the neutral atoms and the radiation field takes the form 
$$\hat{H}_I=-\int d \bx \bff \mu (\bff x) \cdot \nabla \times \bff A(\bff x)$$
where $\bff \mu$ is the magnetization density and $\bff A$ is 
the electromagnetic field at position $\bff x$.
We have shown \cite{BazBar14}, that in effective low energy theory \cite{Kap05}
$\bff \mu$ has contributions from one-body current as well as more body currents 
\be \label{currents}
\bff \mu(\bff x)=\bff \mu^{(1)}(\bff x)+\bff \mu^{(2)}(\bff x)+ \ldots.
\ee
An effective theory has some UV cutoff $\Lambda$, reflecting some short-range 
physics which is ignored. Naive power-counting suggests that each order in this
expansion is suppressed by a factor of $(Q/\Lambda)^3$, 
where $Q$ is the typical momentum of a particle in the system under 
consideration.
The leading order (LO) one-body current is 
\cite{SonLazPar09} 
\be
\bff\mu^{(1)}(\bff x)= \sum_j\bff s_j \mu_1 \delta(\bff x-\bff r_j),  
\ee
where $\mu_1$ is the magnetic moment of a single particle, 
$\bff{r}_j$ and $\bff{s}_j$ are the position and the spin of particle $j$. 
The two-body contribution to the atomic electro-magnetic current enters at 
the next order (N$^2$LO),
or $(Q/\Lambda)^3$, and takes the form \cite{SonLazPar09}
\be\bff\mu^{(2)}(\bff x)= \sum_{i<j}(\bff s_i+\bff s_j) 
\frac{L_2}{\Lambda^3} \delta(\bff x-\frac{\bff r_i+\bff r_j}{2})
       \delta_{\Lambda}(\bff r_i-\bff r_j)\;
\ee
where $L_2=L_2(\Lambda)$ is the coupling constant between the radiation
field and the four boson fields.
The notation $\delta_{\Lambda}(\bff r)$ stands for Dirac's
$\delta$-function smeared over distance $\hbar/\Lambda$.

In the low energy limit $Q\ll\Lambda$, more particles current 
can be ignored due to suppression by additional $(Q/\Lambda)^3$ factor
associated with each extra particle field.

Using box normalization of volume $\Omega$, the electro-magnetic field reads 
$$ \bff A(\bx)=\sum_{\bff k,\zeta} \sqrt{\frac{\hbar}{2\Omega\omega\epsilon_0}}
  \hat e_{\bk \zeta} (a^\dagger_{\bk \zeta}e^{i\bff k \cdot \bff x}+h.c.) $$
where $\zeta=1,2$ are the linear photon polarizations,
$\omega$ is the photon frequency, $\bk$ its momentum, and
$\epsilon_0$ is the vacuum permeability.

For bosonic systems confined in a strong (static) magnetic field,
the initial and final state atomic wave functions can be written as 
a product of spin and configuration space terms, $\Psi=\psi_{L M}\chi_{M_F}$.
The spin wave function $\chi_{M_F}$ is the symmetrized function 
$\chi_{M_F}={\cal S}\left[|m_F(1)\ket|m_F(2)\ket\ldots | m_F(N)\ket\right]$
with magnetic quantum number 
$M_F=\sum_j m_F(j)$.
The spatial wave function 
$\psi_{L M}=\psi_{L M}(\bff{r}_1,\bff{r}_2,\ldots \bff{r}_N)$ is a symmetric
function with angular momentum quantum numbers $L M$.
Using this factorization, the one-body transition matrix element 
in (\ref{golden_rule}) takes the form
\be\eqalign{
\bra f ,\bk \zeta| \hat H_I^{(1b)} | i \ket=
-i \mu_1\sqrt{\frac{\hbar}{2\Omega\omega\epsilon_0}}\times \\  
\sum_{j=1}^N 
\bra \chi^f_{M'_F} | \bff s_j \cdot (\bk \times \hat e_{\bk \zeta}) |
\chi^i_{M_F} \ket 
\bra \psi^f_{ L' M'} | e^{i\bff k \cdot \bff r_j}| \psi^i_{ L  M} \ket,
}\ee
whereas the two-body part reads 
\be\eqalign{\label{hi2b}
\bra f ,\bk \zeta |\hat H_I^{(2b)} | i \ket= 
-i\sqrt{\frac{\hbar}{2\Omega\omega\epsilon_0}} 
   \frac{L_2}{\Lambda^3} \times \\
   \sum_{i<j}^N 
   \bra \chi^f_{M'_F} | (\bff s_i+\bff s_j) \cdot (\bk \times \hat e_{\bk \zeta}) |
   \chi^i_{M_F} \ket \times \\
   \bra \psi^f_{ L' M'}  | e^{i\bk \cdot (\bff r_i+ \bff r_j)/2}
   \delta_{\Lambda}(\bff r_i- \bff r_j)| \psi^i_{ L  M} \ket.
}\ee

Using spherical notation, the spin operator reads
$
\bff s \cdot (\bk \times \hat e_{\bk \zeta})=\sum_\lambda
(-)^{\lambda}s_{-\lambda} \cdot (\bk \times \hat e_{\bk \zeta})_\lambda,
$
where $\lambda=0,\pm 1$.
The geometrical factor, $\eta=(\hat k \times \hat e_{\bk \zeta})_\lambda,$
is maximal (minimal) for the frozen spin $\lambda=0$ case
(spin flip $\lambda=\pm 1$ case),
when the rf magnetic component is parallel to the static magnetic field.

The photon wavelength of rf radiation is much larger than the typical 
dimension of the system $R$, therefore $kR\ll1$ and the lowest order 
in $kR$ dominates the interaction. 
When the photon can induce a Zeeman state change, i.e. spin-flip, the leading 
contribution comes at order $k$. Energy is delivered to the system
through the spin matrix element and we can approximate 
$e^{i\bff k \cdot \bff r}\simeq 1$, to get 
\be\eqalign{
\label{spin-flip}
|\bra f ,\bk \zeta| \hat H_I^{(1b)} | i \ket |^2=
\frac{\mu_1^2\hbar k \eta^2}{2\Omega c \epsilon_0}
\sum_{\lambda=\pm 1} \\ 
|\bra \chi^f_{M'_F} |\sum_{j=1}^N s_{j,\lambda}| \chi^i_{M_F} \ket |^2
|\bra \psi^f_{ L' M'} | \psi^i_{ L M} \ket |^2
 \delta_{L,L'} \delta_{M,M'}.
}\ee
The last term on the rhs of (\ref{spin-flip}) is just the Franck-Condon
factor, that appears in the commonly used theory of photoassociation, 
such as  Eq. (3) in Ref. \cite{ChiJul05}.

When the rf photon cannot induce change in the spin structure of the system, 
$\lambda=0$, and we call the process a frozen spin reaction. 
In this case $M_F'=M_F$, $|\chi^f_{M_F}\ket = | \chi^i_{M_F}\ket $ and
the transition matrix element can be written as
\be\eqalign{ \label{frozen-spin}
\bra f,\bk \zeta | \hat H_I^{(1b)} | i\ket =& -i \mu_1 
\sqrt{\frac{\hbar}{2\Omega\omega\epsilon_0}} k \eta \bra s_{0} \ket \times \\ &
\bra \psi^f_{L' M'} | \sum_{j=1}^N e^{i\bff k \cdot \bff r_j} | \psi^i_{L M} \ket \;,
}\ee
where 
$\bra s_{0} \ket = \frac{1}{N} \sum_j \bra \chi^i_{M_F} |s_{j,0}| \chi^i_{M_F}\ket$ 
is the average single particle magnetic moment, which plays the role 
of an effective charge.

In the long wavelength limit, the exponent can be expanded to yield
\be\eqalign{ \label{expand}
\sum_{j=1}^N e^{i\bff k \cdot \bff r_j} &\approx  N
+i\sum_{j=1}^N \bk \cdot \bff r_j
-\frac{1}{6} \sum_{j=1}^N k^2 r_j^2 \cr &
-\frac{4\pi}{15} \sum_{j=1}^N k^2 r_j^2 \sum_m Y_{2 -m}(\hat{k}) Y_{2 m}(\hat{r}_j),
}\ee
where $Y_{lm}$ are the spherical harmonics.
Clearly this expansion has transparent physical meaning.
The zero order operator is proportional to 1 and stands for elastic 
interaction. In photon emission reaction this process is forbidden by 
energy conservation.
Next comes at first order the dipole. Dealing with identical particles,
this term is proportional to the center of mass and hence cannot affect 
internal degrees of freedom.
Two operators appear at second order:
the $r^2$ operator, corresponding 
to $s$-mode reaction, and the quadrupole terms, corresponding 
to $d$-mode reaction.
Summing over the initial and final magnetic numbers $M$ and $M'$, 
the transition matrix element reads \cite {BazLivBar12}
\be\eqalign{ \label{t-frozen}
\sum_{M,M'}
| \bra f,\bk \zeta | \hat H_I^{(1b)} | i \ket |^2  = 
\frac{4\pi \hbar k^5 \mu_1^2 \eta^2}{2\Omega c \epsilon_0}
 \ev{s_0} \\ 
\left(
\frac{1}{6^2} 
| \bra \psi^f_{L'} \Vert \sum_{j=1}^N r_j^2 Y_0\Vert \psi^i_{L} \ket |^2
+\frac{1}{15^2} 
| \bra \psi^f_{L'} \Vert \sum_{j=1}^N r_j^2 Y_2(\hat r_j)  \Vert \psi^i_{L} \ket |^2
\right).}
\ee

For the two body current, we can again utilize the long wave approximation
$e^{i\bk \cdot (\bff r_i+ \bff r_j)/2} \approx 1$ to get 
\be\eqalign{ \label{2-body}
|\bra f ,\bk \zeta |\hat H_I^{(2b)} | i \ket|^2= 
\frac{4 \hbar k}{2\Omega c \epsilon_0}
\frac{L_2^2 \eta^2}{\Lambda^6} \ev{s_0} \times \\
| \bra \psi^f_{ L' M'} | \sum_{i<j}^N \delta_{\Lambda}(\bff r_i- \bff r_j)| \psi^i_{ L  M} \ket |^2. }\ee

To find the most relevant operators for the photoassociation process
one should consider both the low energy 
expansion (\ref{currents}) and the long wavelength expansion (\ref{expand}).
Dealing with trimer photoassociation, the photon energy has to be of the order 
of the trimer binding energy $E_3$. Therefore the long wavelength expansion 
parameter can be written as 
$k r \approx \sqrt{E_3/Mc^2}\approx Q/Mc,$
where we used $r\approx \sqrt{\hbar^2/M E_3}$.
Therefore one should compare the long wavelength expansion parameter,
$k r \approx Q/Mc$ to the low energy expansion parameter $Q/\Lambda$.
If $Q/\Lambda \ll {\Lambda}/Mc$, the two-body currents appearing at order
$(Q/\Lambda)^3$ are much smaller than the second order 
$(kr)^2\approx (Q/Mc)^2$ terms. This \emph{normal hierarchy} case is the 
situation in the limit $Q \longrightarrow 0$. 
In the other extreme, when $\Lambda \ll Mc$, the two-body current is more 
important than the one-body current, proportional to $(Q/Mc)^2$.
This case of \emph{strong hierarchy} is typical for photoassociation
experiments in ultracold atoms \cite{LomOttSer10,NakHorMuk11,MacShoGro12}.
There, the separation between the mass scale and the binding energy is much 
bigger than the ratio between the scattering length and the effective range.

\section{The Three Body Problem}

The Schroedinger equation governs the dynamics of a quantum 3 particle 
system
\be \label{Hamiltonian}
\left(T+W\right)\psi=E\psi,
\ee
where $T$ is the kinetic energy operator in the center of mass frame 
and $W$ is the potential. Here we assume a short range 2-body forces, 
thus $W=\sum_{i<j}V(|\bff{r}_i-\bff{r}_j|)$.
To solve this problem we use the hyperspherical coordinates
and the adiabatic expansion \cite{Mac68,EsrGreBur99}. 
In the following section we review these methods and extend them for
the $L>0$ case. 

\subsection{The Hyperspherical coordinates}
To eliminate the center of mass motion, we define the Jacobi coordinates,
$$
\bff x_i=\sqrt \frac {1}{2} (\bff r_j-\bff r_k),\quad \bff y_i = 
\sqrt \frac {2}{3}\left(-\bff r_i + \frac{\bff r_j+\bff r_k}{2}\right)
$$
where $\{ijk\}$ is a cyclic permutation of $\{123\}$.
The relation between two sets of Jacobi coordinates $(\bff x_i, \bff y_i)$
and $(\bff x_j, \bff y_j)$ is given by the kinematic rotation,
\be \label{KinRot}
\left ( \begin{array}{c} \bff x_j \\ \bff y_j \end{array}  \right ) =
\left ( \begin{array}{cc} -\cos \phi_{ij} & \sin \phi_{ij} \\ -\sin \phi_{ij} & 
-\cos \phi_{ij}\end{array}  \right )
\left ( \begin{array}{c} \bff x_i \\ \bff y_i \end{array}  \right ).
\ee
For the case of 3 identical particles, $\phi_{ij}=\pm \pi/3$, where the 
sign is fixed by the parity of the permutation $\{ijk\}$.
For each set of Jacobi coordinates we can define a set of 
hyperspherical coordinates $(\rho,\Omega_i)$,
where $\rho^2=x_i^2+y_i^2,$ $\Omega_i=(\alpha_i,\hat x_i, \hat y_i)$, and 
$\tan \alpha_i=x_i/y_i.$
Two different sets of hyperspherical coordinates are related through 
\be \label{alpha}
2\sin^2\alpha_j=1-\cos 2\phi_{ij} \cos 2\alpha_i 
- \gamma_i \sin 2\phi_{ij} \sin 2\alpha_i,
\ee
where $\gamma_i=\hat x_i \cdot \hat y_i.$
The hyperradius $\rho$ is invariant under kinematic rotations, particle
permutations, and  
therefore under change of coordinates sets.

In the hyperspherical coordinates the kinetic energy operator reads,
\be
T=-\frac{\hbar^2}{2m} \left (\frac{\partial^2}{\partial\rho^2} 
+\frac{5}{\rho}\frac{\partial}{\partial\rho}-\frac{\hat K ^2}{\rho^2}
\right).
\ee
Here, $\hat K^2$ is the square of the grand angular momentum operator,
\be
\hat K^2=-\frac{1}{\sin 2\alpha}\frac{\partial^2}{\partial \alpha^2}\sin2\alpha
+\frac{\hat l^2_x}{\sin^2\alpha}+\frac{\hat l^2_y}{\cos^2\alpha}-4\;,
\ee
and $\hat l_x$, $\hat l_y$ are the angular momentum operators corresponding 
to $\bff x,\bff y$ coordinates. 
The hyperspherical presentation of a central two-body potential reads,
\be
\sum_{i<j}V(|\bff{r}_i-\bff{r}_j|)=\sum_i V(\sqrt 2 \rho \sin \alpha_i).
\ee

\subsection{The Adiabatic Expansion}

Next we apply the adiabatic expansion \cite{Mac68} and write 
the wave function $\psi$ in the form
\be
\psi(\rho,\Omega)=\sum_n \rho^{-5/2}\mathcal R_n(\rho)\Phi_n(\rho,\Omega).
\ee
The hyperspherical functions $\Phi_n(\rho,\Omega)$ are the solutions of the 
hyperangular equation,
\be\eqalign{
 \left ( \hat K ^2 + \frac{2m\rho^2}{\hbar^2}\sum_i 
V(\sqrt 2 \rho \sin \alpha_i) +4 \right)\Phi_n(\rho,\Omega)= \cr
 \nu_n^2 \Phi_n(\rho,\Omega), }\ee
corresponding to the eigenvalue $\nu_n^2$.
The hyperradial functions $\mathcal R_n(\rho)$ are the solutions 
of the hyperradial equation,
\be\eqalign{\label{hre} \left (-\frac{\partial^2}{\partial\rho^2}+V_{\mathrm{eff}}(\rho)
-\epsilon\right)\mathcal R_n(\rho)= \cr
\sum_{n'; n'\ne n}(2P_{n n'}\frac{\partial}{\partial\rho}
+Q_{n n'})\mathcal R_{n'}(\rho) }\ee
where $\epsilon=2mE/\hbar^2$, $V_{\rm{eff}}$ is the effective hyperradial 
potential, and $P_{n n'}, Q_{n n'}$ are the non-adiabatic couplings.
The effective potential is given by
$$ V_{\mathrm{eff}}(\rho)=\frac{\nu_n^2(\rho)-1/4}{\rho^2}-Q_{n n}, $$
and the non-adiabatic couplings are 
\be\eqalign{\label{nac}
 P_{nn'}(\rho)=\left \bra \Phi_n(\rho,\Omega) \left \vert \frac{\partial}{\partial\rho} \right \vert \Phi_{n'}(\rho,\Omega)\right \ket_\Omega \cr
 Q_{nn'}(\rho)=\left \bra \Phi_n(\rho,\Omega) \left \vert \frac{\partial^2}{\partial\rho^2}\right \vert \Phi_{n'}(\rho,\Omega)\right \ket_\Omega.}
\ee
The expectation value $\bra ... \ket_\Omega$ stands for integration 
over the hyperangles $\Omega$.

\subsection{The Hyperangular Equation}

For low energy physics, when the extension of the wave function is
much larger than the range of the potential, one can utilize the zero range
approximation. In this approximation the lateral extension of the potential is
neglected all together, and the action of the potential is represented through
the appropriate boundary conditions.
For a two-particle system the low energy interaction is dominated by the
$s$-wave scattering length $a$ and the wave function fulfills the boundary
condition $[u'/u]_{r=0}=-1/a$. The corresponding 3-body condition is
\be\label{bc}
\left[\frac{1}{2\alpha_i \Phi}\frac{\partial}{\partial \alpha_i}2 \alpha_i \Phi
\right]_{\alpha_i=0}=-\sqrt 2 \frac{\rho}{a}\;.
\ee

Dealing with wave functions of definite total angular momentum 
quantum numbers $L,M$, we use a Faddeev-like decomposition,
\be \label{FD}
\Phi(\rho,\Omega)=N_i \sum_i \sum_{l_x,l_y} \phi_i^{l_x,l_y}(\rho,\alpha_i)
Y_{l_x,l_y}^{L,M}(\hat x_i,\hat y_i),
\ee
where $N_i$ is a normalization constant and 
\be
Y_{l_x,l_y}^{L,M}(\hat x,\hat y)=\sum_{m_x,m_y}\bra l_x m_x l_y m_y | L M \ket
Y_{l_x m_x}(\hat x)Y_{l_y m_y}(\hat y).
\ee
It should be noted that for a bosonic system $l_x$ must be even.

In the zero range approximation each Faddeev component is a solution of
the free hyperangular equation, 
\be \label{haf}
(\hat K^2 - \Lambda ) \phi_i(\rho,\Omega_i)Y_{l_x,l_y}^{L,M}(\hat x,\hat y) = 0\;,
\ee
with the appropriate boundary conditions.

One family of solutions to (\ref{haf}) is the hyperspherical 
harmonics \cite{Ave93},
\be \label{hsh}
\mathcal Y^{KLM}_{l_xl_y}(\Omega)=\varphi^K_{l_x,l_y}(\alpha)
Y_{l_x,l_y}^{L,M}(\hat x,\hat y)
\ee
where
\be
\varphi^K_{l_x,l_y}(\alpha)=
N^K_{l_x,l_y}\sin^{l_x}\alpha\cos^{l_y}\alpha P^{(l_x+1/2,l_y+1/2)}_n(\cos 2\alpha).
\ee
Here $P_n^{(\alpha\beta)}$ are the Jacobi polynomials, $n=(K-l_x-l_y)/2$, 
and 
$$
N^K_{l_x,l_y}=
\sqrt{\frac{2n!(K+2)(n+l_x+l_y+1)!}{\Gamma(n+l_x+3/2)\Gamma(n+l_y+3/2)}}.
$$
Note that $n$ is a non-negative integer and therefore $K \geq l_x+l_y$.
The hyperspherical harmonics correspond to $\Lambda=K(K+4)$ and 
are regular at $\alpha=0$ and $\alpha=\pi/2$.

Another family of solutions can be found by generalizing (\ref{hsh}) 
to a non-integer $K$ (and $n$), 
loosing the regularization in one edge.
Due to the boundary condition (\ref{bc}), we 
need such solution for the interaction channel, i.e. the $l_x=0$ component.
Such solutions, corresponding to the 
eigenvalue $\Lambda=\nu^2-4$, are
$P_\nu^L(\alpha)/\sin 2\alpha$ and $Q_\nu^L(\alpha)/\sin 2\alpha$
\cite{CobFedJen97}, where 
\be
P_\nu^L(\alpha)=\cos^L\alpha
\left(\frac{\partial}{\partial\alpha}\frac{1}{\cos\alpha}\right)^L
\sin\left(\nu\left(\alpha-\frac{\pi}{2}\right)\right)
\ee
and 
\be
Q_\nu^L(\alpha)=\cos^L\alpha
\left(\frac{\partial}{\partial\alpha}\frac{1}{\cos\alpha}\right)^L
\sin\left(\nu \alpha\right).
\ee
The first (second) solution is regular at $\alpha=\pi/2$ ($\alpha=0$) 
and not regular at $\alpha=0$ ($\alpha=\pi/2$).

For evaluating the rf transition matrix elements we need to consider 
$L=0$ and $L=2$ states. Their explicit form is given by
\be\eqalign{\label{PQ0}
P_\nu^0(\alpha)= \sin\left(\nu\left(\alpha-\frac{\pi}{2}\right)\right) \\
Q_\nu^0(\alpha)= \sin (\nu \alpha),
}\ee
and
\be\eqalign{\label{PQ2}
P_\nu^2(\alpha)&=3 \nu \tan \alpha
\cos\left(\nu\left(\alpha-\frac{\pi}{2}\right)\right) \\
&-(2 +\nu^2 - 3 \sec ^2 \alpha)
 \sin \left(\nu\left(\alpha-\frac{\pi}{2}\right)\right)\\
Q_\nu^2(\alpha)&=3 \nu \tan \alpha \cos (\nu \alpha) -(2 + \nu^2 - 3 \sec ^2
\alpha) \sin (\nu \alpha).\nonumber }
\ee

\subsection{The Projection Operator}
It is convenient to project the wave function (\ref{FD}) on a single
coordinate system  
\cite{CobFedJen97}. The resulting expression takes the form
\be\eqalign{ \label{Sol}
  \Phi(\Omega_i)&=N_i \sum_{l_x,l_y} \left\{ \phi_i^{l_x,l_y}(\alpha_i) \right.\\
  +\sum_{j \ne i} &\sum_{l_x',l_y'} 
  R_{ij}^{(l_xl_y)(l_x'l_y')}\left[\phi_j^{l_x',l_y'}(\alpha_j)\right](\alpha_i) 
  \left. \vphantom {\phi_i^{l_x}} \right\}
  Y_{l_x,l_y}^{L,M}(\hat x_i,\hat y_i)\;,}
\ee
where $R_{ij}$, the projection operator is given by
\be\eqalign{\label{PO}
R_{ij}^{(l_xl_y)(l_x'l_y')}&[\phi(\alpha_j)](\alpha_i) \equiv  \\ 
\int d\hat x_i \int d\hat y_i &Y_{l_xl_y}^{LM*}(\hat x_i,\hat y_i)
\phi(\alpha_j)Y_{l_x'l_y'}^{LM}(\hat x_j,\hat y_j)}
\ee

To calculate $R_{ij}$ we first study the limit $\alpha_i=0$.
From (\ref{KinRot}) it follows that in this case,
$\bff x_j=\pm \sqrt 3 \bff y_i/2 $, 
$\bff y_j=-\bff y_i/2$ and $\alpha_j=\pi/3$.
The $\pm$ sign depends on the direction of the kinematic rotation.
The integral in (\ref{PO}) can now be evaluated, yielding 
\be\eqalign{ \label{Rot}
&R_{ij}^{(l_xl_y)(l_x'l_y')}[\phi(\alpha_j)](\alpha_i=0) = \\
&\phi(\pi/3)(\mp)^{l'_x}\delta_{l_x,0}\delta_{l_y,L}
\sqrt{(2l'_x+1)(2l'_y+1)}\threej {l_x'}{l_y'}{L}{0}{0}{0},
}\ee
where the last term is the 3-j symbol.
Some conclusions emerge: 

a. The projection at $\alpha_i=0$ is zero for all waves but $l_x=0,l_y=L$; 
   similar considerations show that the projection at 
   $\alpha_i=\pi/2$ is zero for all waves but $l_x=L,l_y=0$.

b. Due to the $(\mp)^{l'_x}$ factor, for odd $l'_x$ 
   the two rotations cancel each other.

c. The 3-j symbols equal zero unless $l_x'+l_y'+L$ is even. 

d. Using this result, an explicit formula for the $l_x=0$ or $l_y=0$ 
   Raynal-Revai coefficients $\bra l_x l_y | l'_x l'_y \ket_{KL}$ 
   \cite{RayRev70} can be constructed. See \ref{RayRev} for details.

Focusing on the $l_x=0$, $l_y=L$ case, the projection operator 
can be computed for any $\alpha_i$ \cite{NieFedJen01}. 
This is because for $\alpha_i\le\pi/3$ the rotated function is an eigenfunction 
of the kinetic energy operator with the same eigenvalue, but is regular 
for $\alpha_i=0$, therefore it is proportional to $Q_\nu^L$. The proportionality
constant can be calculated by matching with the known value at $\alpha_i=0$. 
Similarly, the rotation for $\alpha_i\ge\pi/3$ is proportional to $P_\nu^L$, 
and the proportionality constant can be found by matching at $\alpha_i=\pi/3$.

Finally, defining 
$\tilde P_\nu^L=P_\nu^L/\sin(2\alpha)$ and $\tilde Q_\nu^L=Q_\nu^L/\sin(2\alpha)$,
\be\eqalign{
R_{ij}^{(0L)(0L)}[\tilde P_\nu^L(\alpha_j)](\alpha_i)=(-1)^L\times \\ \nonumber
\cases{
\tilde {Q}_{\nu}^L(\alpha_i) \tilde {P}_{\nu}^L(\pi/3) / \tilde {Q}_{\nu}^L(0)
& $0 \le \alpha_{i} \le \pi /3 $\\
\tilde {P}_{\nu}^L(\alpha_i) \tilde {Q}_{\nu}^L(\pi/3) / \tilde {Q}_{\nu}^L(0)
& $\pi/3 \le \alpha_i \le \pi/2.$}}
\ee

\subsection{Imposing the Boundary Condition}

In the limit of infinite scattering length, the adiabatic expansion
decouples and the non-diagonal couplings $P_{nn'}, Q_{nn'}$ vanish 
\cite{CobFedJen97}.
In this limit, the lowest energy hyperangular function takes the form
\be
  \phi_i(\alpha_i)=P_\nu^L(\alpha_i)/\sin(2\alpha_i)\;.
\ee
The eigenvalues $\nu$ will emerge as we impose the 
appropriate boundary conditions at $\alpha_i=0$.
We have seen in (\ref{Rot}), that the rotation $R_{ij}$ at 
$\alpha_i=0$ includes only the $l_x=0$ partial wave, therefore 
for small $\alpha_i$ (\ref{Sol}) is simply given by \cite{FedJen02},
\be\eqalign{
  \frac{\sin 2\alpha_i}{N_i} \Phi(\Omega_i)= &P_\nu^L(\alpha_i)+
  2(1+(-)^L)\alpha_i \tilde P_\nu^L(\pi/3) \\ &+ O(\alpha_i^2)}
\ee
Therefore the boundary conditions (\ref{bc}), turn into 
\be
  \frac{\partial P_\nu^{L}(0)}{\partial \alpha}+\frac{4}{\sqrt 3}
  (1+(-)^L)P_\nu^{L}(\pi/3)=-\frac{\sqrt 2 \rho}{a} P_\nu^{L}(0) \;.
\ee

For $L=0$ the resulting equation for $\nu$ is \cite{Efi71}
\be\label{l0}
  \nu\cos(\nu\pi/2)-\frac{8}{\sqrt 3}\sin (\nu\pi/6)=\frac{\sqrt 2 \rho}{a}
  \sin(\nu\pi/2).
\ee
For $|a|=\infty$ the {solution with lowest $\nu^2$ is} $\nu_0\approx
1.00624i$, corresponding to the Efimov trimer. For $a > 0$ this solution 
approaches asymptotically a particle scattering from a universal dimer, 
where $V_{\mathrm{eff}}(\rho\longrightarrow\infty)=-1/a^2$ \cite{FedJen01}. 
The spurious solution $\nu=4$ is just $\Phi=0$.

For $L=2$ the corresponding equation for $\nu$ is,
\be\eqalign{\label{l2}
  &\nu(4-\nu^2)\cos(\nu\pi/2)+
  24 \nu \cos(\nu \pi/6) + \\ 
  &\frac{8}{\sqrt3} (\nu^2-10) \sin (\nu \pi/6)=
  -\frac{\rho}{a}(\nu^2-1)\sin(\nu\pi/2)}
\ee
For $|a|=\infty$ the lowest non-trivial solution is
$\nu_2\approx2.82334$. 
For $a > 0$ this solution asymptotically converges to particle-dimer
$d$-wave scattering, where 
$V_{\mathrm{eff}}(\rho\longrightarrow\infty)=L(L+1)/\rho^2-1/a^2$.
The $\nu=0,1,2$ correspond to the spurious solution $\Phi=0$.
For $\rho \gg |a|$ solutions with $\nu$ an even integer, are just 
the regular, free, hyperspherical harmonics.
 
\begin{figure}
\includegraphics[width=8.6 cm]{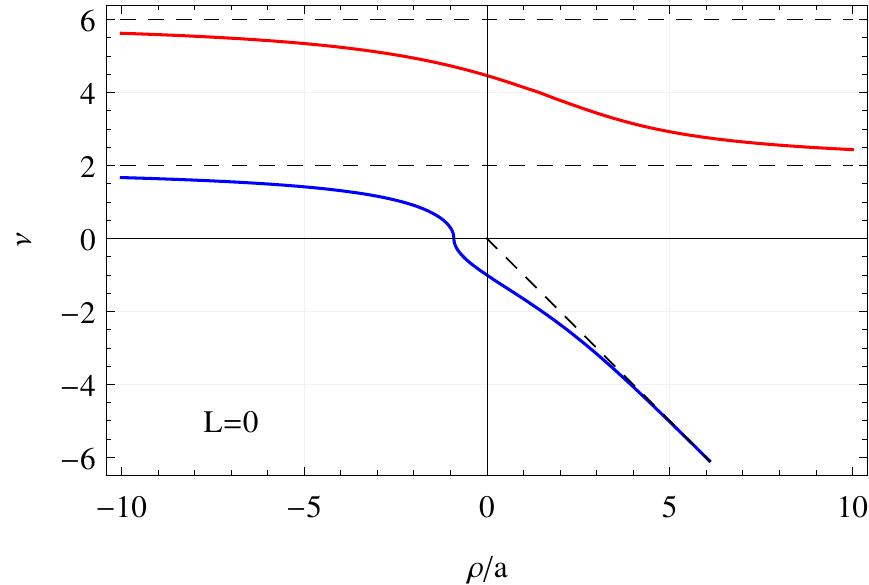}
\includegraphics[width=8.6 cm]{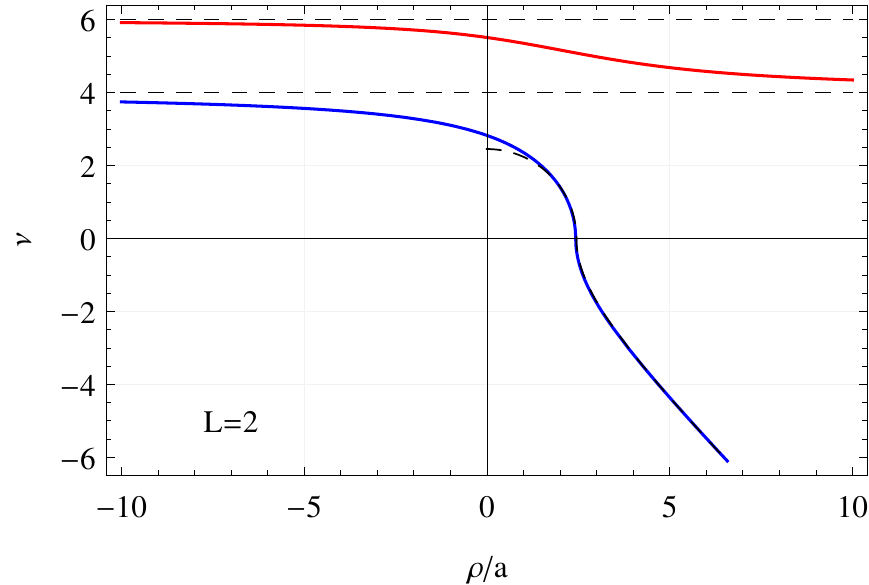}
\caption{
  \label{Fig:L2} (Color online) The two lowest angular eigenvalue $\nu$ as a
  function of $\rho/a$ for $L=0$ (Upper panel) and $L=2$ (Lower panel). 
  The asymptotic solutions are dashed. 
  Negative $\nu$ corresponding to imaginary solution.}
\end{figure}

The two lowest eigenvalues of the hyperangular equation, with the appropriate
boundary conditions (\ref{l0}),(\ref{l2}), are plotted in Fig. \ref{Fig:L2}
for $L=0$ and $L=2$, with their asymptotic forms.

\subsection{{The Hyperradial Equation}}

To proceed analytically, we focus on the unitary limit, $|a|\rightarrow\infty$. 
In this case $\nu_L(\rho)=\nu_L$ and $\Phi(\rho,\Omega)=\Phi(\Omega)$, 
therefore the non-adiabatic couplings, Eq. (\ref{nac}), vanish and 
the hyperradial equation for $ \mathcal R(\rho)/\sqrt \rho$ is just the 
Bessel equation, 
\be\label{Bessel}
-\frac{\partial^2 \mathcal R_n(\rho)}{\partial \rho^2} +\frac{\nu_L^2-1/4}{\rho^2} \mathcal R_n(\rho)
=\epsilon \mathcal R_n(\rho).
\ee
In order to calculate the trimer photoassociation, caused by $s-$ and $d$-wave 
operators, we have to solve the $L=0$ bound state as well as the 
$L=0,2$ continuum states.
The needed solutions are the followings:

I. A bound state ($\epsilon<0$) with $L=0$, and $\nu_0\approx1.00624i$.
In this case the relevant solution is proportional to the modified 
Bessel function of the second kind and imaginary order,
$\sqrt \rho K_{\nu_0}(\kappa \rho)$, 
where $\kappa=\sqrt {-\epsilon}$. 
At the origin, this solution behaves like 
$\sin (\nu \ln (\kappa \rho/2)-\gamma_\nu)$, 
where $\gamma_{\nu_0}\approx -0.301$. 
Therefore regularization is needed to avoid collapse, 
e.g. setting $\mathcal R(\rho \le \rho_0)=0$ for some finite $\rho_0$.
The result is the discrete Efimov spectrum,
\be
\frac{\epsilon_m}{\epsilon_0}=e^{-2\pi m/|\nu_0|}\approx 515^{-m}.
\ee
The normalized wave functions are 
\be \label{bound_wf}
\mathcal R^{(m)}_B(\rho)=N_B \kappa_m \sqrt \rho K_{\nu_0}(\kappa_m \rho)
\ee
where 
$N_B=\sqrt{2 \sin \nu_0 \pi/\nu_0 \pi} \approx 2.730$,
and 
\be
\frac{\kappa_m}{\kappa_0}=e^{-\pi m/|\nu_0|}\approx 22.7^{-m}.
\ee

II. Scattering state ($\epsilon>0$) with $L=0$, and $\nu_0\approx1.00624i$.
The solution is composed of the real part of the Bessel functions of the
first and second kind of imaginary order,
\be \label{cont_wf1}
\mathcal R_s(\rho)=
\sqrt{\frac{q \rho}{2 R}} \left[ \sin \delta \mathrm{Re}[J_{\nu_0}(q \rho)] +
\cos \delta \mathrm{Re}[Y_{\nu_0}(q \rho)] \right]
\ee
where $q=\sqrt {\epsilon}$ and we assume normalization in a sphere of 
radius $R$.
The phase shift $\delta$ is to be found from the boundary condition, 
$\mathcal R_s(\rho_0)=0$.

In the scattering problem, functions with higher $\nu$, corresponding to
the same angular momentum $L=0$ and energy $\epsilon$, are also legitimate 
solutions. However, due to the orthogonality in $\Omega$, there is no overlap 
between these functions and the bound state, and therefore no contribution 
to the photoassociation transition matrix elements.

III. Scattering state ($\epsilon>0$) with $L=2$, and real $\nu_2$. 
The solutions are composed of the Bessel functions of first and second kind,
\be \label{cont_wf2}
\mathcal R_d(\rho)=\sqrt{\frac{\pi q \rho}{R}}
\left[ \sin \delta_d J_{\nu_2}(q \rho) + \cos \delta_d Y_{\nu_2}(q \rho) \right]\;.
\ee
The phase shift $\delta_d$ is to be 
determined by the condition $\mathcal R_d(\rho_0)=0$.
The three lowest values for $\nu_2$ are, 
$\nu_2 \approx 2.823$, $\nu_2^1 \approx 5.508$ and $\nu_2^2 \approx 6.449$.

\section{The Transition Matrix Elements}

Now that we have solved the Schroedinger equation and obtained the 
$L=0$ bound state wave function and the $L=0,2$ scattering wave functions,
we are in a position to evaluate the transition matrix elements, 
in both hierarchies.
For the \emph{normal hierarchy} case the needed matrix elements (\ref{t-frozen})
read,
\be
I_{1b}^s=\bra \psi^f_{L'} \Vert \sum_{j=1}^N r_j^2 Y_0 \Vert \psi^i_L \ket,
\ee
for the $s-$mode transition and
\be
I_{1b}^d=\bra \psi^f_{L'} \Vert \sum_{j=1}^N r_j^2 Y_2(\hat r_j) \Vert \psi^i_L \ket
\ee
for the $d-$mode.
For the \emph{strong hierarchy} case (\ref{2-body}) the matrix element is,
\be
I_{2b}=\bra \psi^f_{L'} \Vert \sum_{i<j}^N \delta_{\Lambda}(\bff r_i- \bff r_j)
      \Vert \psi^i_L \ket,
\ee
These matrix elements are proportional to integrals of the type
\be \label{IJm}
    {\cal I}_{J}(\nu,m) = \int_{\rho_0}^\infty d\rho\, K_{\nu_0}(\kappa \rho)
    \rho^{m+1} \mathrm{Re}[J_{\nu}(q \rho)].
\ee
or ${\cal I}_{Y}(\nu,m)$ with $Y_{\nu}$ replacing $J_{\nu}$.

Taking the lower limit to zero, these integrals can be evaluated analytically,
\be \label{IJmx}
 {\cal I}_J(\nu,m) \approx {\rm Re}\left[
\frac{2^m N_{\nu_0,\nu}^m}{\kappa^{m+2}}\left(\frac {q}{\kappa}\right)^\nu
{}_2F_1\left(a,b;c;-(q/\kappa)^2\right)\right]
\ee
where $_2F_1$ is the hypergeometric function with parameters
$a= \frac{m-\nu_0+\nu}{2}+1$, $ b=\frac{m+\nu_0+\nu}{2}+1$, and $c=\nu+1$,
and $N_{\nu_0,\nu}^m=\Gamma(a)\Gamma(b)/\Gamma(c)$. 
For $q < \kappa$ this approximation amounts to an error 
$\Delta{\cal I}/{\cal I} \sim x_n^{m+2}$, where $x_0\approx 0.06$ is the 
largest zero of $K_{\nu_0}(x)$ and $x_n$ is the $n$'th root.
Similar results can be obtained for ${\cal I}_{Y}$.

\subsection{Transition matrix elements in the normal hierarchy case}

\emph{The $r^2$ matrix element.}
The $r^2$ operator connects the $L=0$ bound state to the $L=0$ scattering 
state. We note that $\sum_j r_j^2 = \rho^2 + 3R_{CM}^2$
where $\bff R_{CM}$ is the center of mass coordinate. As 
the CM cannot induce inelastic transition, the matrix element is reduced 
into the hyperradial integral 
\be \label{hr_int} 
I_s(\kappa,q)=\int_0^\infty d\rho 
\mathcal R_B^*(\rho) \rho^2 \mathcal R_s(\rho).
\ee
Substituting Eqs. (\ref{bound_wf}) and (\ref{cont_wf1}) into (\ref{hr_int}), 
we get 
\be  \label{Is2}
I_s(\kappa,q)=\sqrt{\frac{q}{2R}} N_B \kappa \left[ I_J(\nu_0,2) \sin \delta + 
I_Y(\nu_0,2) \cos \delta \right ].
\ee
This integral depends on the momentum of the bound state $\kappa$ and 
the scattering state $q$. The maximal value of $I_s(\kappa,q)$ is
obtained when $q\approx\kappa/2$. 
For higher Efimov states, this integral scales such as 
$$p^{5/2} I_s(\kappa_m,q p)=I_s(\kappa_0,q)$$
where $p=\kappa_m/\kappa_0=22.7^{-m}$, 

It should be noted that (\ref{hr_int}) leads to an oscillatory
log-periodic response function, see discussion in \cite{BazBar14}.
At threshold, the matrix element
(\ref{hr_int}) gets a particularly simple form 
which can be well approximated by \cite{BazBar14}
\be
I_s(\kappa,q) \approx C \kappa^{-3}\sqrt{q}\left(1+\frac{B_3}{2}
\cos(2|\nu_0|\ln\frac{q}{\kappa})\right)
\ee
where $C$ is a constant that contains the normalization factors, and 
$B_3 \approx 8.475\%$ is the normalized amplitude of these oscillations.  
These oscillations modulate the matrix element all the way to the high energy 
tail. Whereas log periodic oscillations at the high energy tail appear in all
partial waves \cite{BraKanPla11}, at threshold the oscillations appear only in 
this $s$-wave transition.

\emph{The quadrupole matrix element.} 
The quadrupole operator, $\sum_j r_j^2 Y_2^M(\hat r_j)$ connects the $L=0$ 
bound state with $L=2$ scattering states. Using
$ \bff r_j=\bff R_{CM} - \sqrt\frac{2}{3} \bff y_j$
and working in center of mass coordinate system,
this operator reads $\frac{2}{3}\sum_j y_j^2 Y_2^M(\hat y_j)$. 
Setting  $y_j^2=\rho^2 \cos ^2 \alpha_j$, the reduced matrix element 
can be written as
\be
\bra \psi_B \Vert \sum_j r_j^2 Y_2 \Vert \psi_d \ket =
\frac{3}{2}\sqrt{5}I_\rho I_\Omega.
\ee
Here the hyperradial integral $I_\rho$ is
\be\label{Iq}
I_\rho=\int_0^\infty d\rho \mathcal R_B^*(\rho) \rho^2 \mathcal R_d(\rho).
\ee
Substituting Eqs. (\ref{bound_wf}) and (\ref{cont_wf2}) into (\ref{Iq}),
we get an expression similar to (\ref{Is2}),
\be 
I_\rho(\kappa,q)=\sqrt{\frac{\pi q}{R}} N_B \kappa 
\left[ I_J(\nu_2,2) \sin \delta_d + I_Y(\nu_2,2) \cos \delta_d \right ].
\ee
The hyperangular integral $I_\Omega$, reads
\be
I_\Omega=\int d\Omega \Phi_f(\Omega) \sum_j \cos^2 \alpha_j Y_{20}(\hat y_j)
\Phi_i(\Omega).
\ee
To evaluate this integral, 
the hyperangular wave function, written in mixed coordinate systems,
$$
\Phi_L(\Omega)=\sum_i P_\nu^L(\alpha_i) Y_{0L}(\hat x_i,\hat y_i)/\sin 2\alpha_i
$$ 
is to be integrated in a single coordinate system. 
Transforming to body-fixed coordinate 
system and integrating over the Euler angles, the integral over the 5 
angles ($\alpha_1,\hat x_1,\hat y_1$) is transformed into an integral over 
($\alpha_1,\gamma_1$) \cite{BarEfrLei04}, that is evaluated numerically.
See \ref{Rotation} for details.
For the lowest $\nu_2$ the integral yields $I_\Omega \approx 0.368.$
For the next two $\nu_2^i$'s the integrals are respectively
$0.023$ and $6.9\times10^{-4}$,
therefore these modes give negligible contribution to the reaction rates.

Here as in the previous case, the hyperradial integral $I_\rho$ leads to
log-periodic oscillations in the high energy tail \cite{BazBar14}.

\begin{figure}
\includegraphics[width=8.6cm]{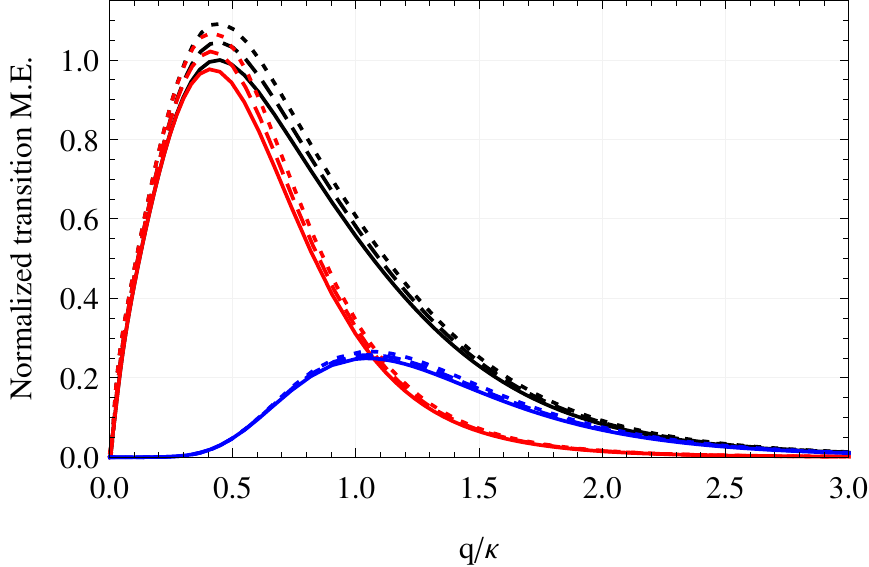}
\caption{\label{ME} (Color online) 
One-body current: The transition matrix element $I_{1b}^2$, 
as a function of the relative momentum $q/\kappa$,
normalized to 1 for the maximum value at $|a|=\infty$
The two modes are presented - $r^2$ (red, peaked around $q = \kappa/2$), and 
quadrupole (blue, peaked around $q = \kappa$),
with their sum (black). 
Shown are the results for the unitary point (solid lines) and 
for finite scattering length, 
$\kappa a=-100$ (dotted lines) and $\kappa a=-50$ (dashed lines).}
\end{figure}

\emph{Relative importance.}
The relative contribution of the $s,d$ modes to the trimer formation is 
displayed in  Fig. \ref{ME}, where the last term in parenthesis on 
the rhs of (\ref{t-frozen}) is presented normalized, along with the 
$s$ and $d$ components. 
Similarly to the dimer formation case \cite{BazLivBar12}, 
the $s$-wave association is peaked around $q\cong\kappa /2$, 
while the $d$-wave association is peaked around $q\cong\kappa$. 

In case of large but finite scattering length,
the eigenvalue of the hyperangular equation, (\ref{l0}) and (\ref{l2}),
depends on $\rho$ and therefore the hyperradial equation (\ref{hre})
is to be solved numerically.
In addition, the non-adiabatic couplings (\ref{nac}) must be considered.
The numerical results for finite $a$ are also shown in Fig. \ref{ME} 
for $\kappa a=-100$ and $\kappa a=-50$.
From the figure it can be seen that the $|a|\longrightarrow \infty$ limit 
is well reproduced if $\kappa |a|> 100$.

\subsection{Transition matrix elements in the strong hierarchy case}
Moving now to study the two-body current, there is no
center of mass contribution, and the transition operator is a scalar exciting 
an $L=0$ bound state into an $L=0$ scattering state.
The transition matrix element is given by,
\be
I_{2b}=
\bra\psi_B|\sum_{i<j}\delta_{\Lambda}(\bff r_i-\bff r_j)|\psi_i\ket=
\frac{3}{\sqrt 8}\bra\psi_B|\delta_{\Lambda}(\bff x)|\psi_i\ket \;.
\ee 
Assuming a simple regularization, 
$\delta_\Lambda(\bff x)=3 \Lambda^3 \Theta(1 - \Lambda x)/4 \pi$,
this matrix element can be calculated,
\be\eqalign{\label{I2b}
I_{2b} = &
\frac{9 \Lambda^2}{64 \pi} 
N_B N_i^2 \kappa \sqrt{\frac{q}{R}}
| \sin (\nu_0 \pi/2) |^2 \times \\
& \left[ I_J(\nu_0,-1) \sin \delta + I_Y(\nu_0,-1) \cos \delta \right ].
}\ee

To renormalize the transition matrix element and remove the cutoff dependence 
we turn to the dimer photoassociation \cite{BazLivBar12}. 
In the zero range approximation the dimer $s$-wave continuum  wave function is 
just $\psi_i=\sqrt{2/R}\sin(q r + \delta)/r$,
and the corresponding bound state wave function is 
$\psi_B=\sqrt{2\kappa} e^{-\kappa r}/r$. Consequently the transition 
matrix element due to the 2-body process is
\be\eqalign{
I_{2b,d} = &
\bra\psi_B|\delta_{\Lambda}(\bff r)|\psi_i\ket = 
\Lambda^3 \frac{6}{4\pi}\sqrt{\frac{\kappa}{R}} 
\biggl(
\frac{q \cos \delta + \kappa \sin \delta}{q^2+\kappa^2} 
\\ 
& - e^{-\kappa/\Lambda}
\frac{q \cos (\delta+q/\Lambda) + \kappa \sin (\delta+q/\Lambda)}
{q^2+\kappa^2} \biggr)
}\ee
which for large values of $\Lambda$ can be approximated as
\be\label{i2bd}
I_{2b,d} = \Lambda^2\frac{6}{4\pi} \sqrt{\frac{\kappa}{R}} \sin \delta +
O(\Lambda)
\ee
Comparing this result to (\ref{I2b}), one can see that the cutoff dependence
of the 2-body current operator, $L_2=L_2(\Lambda)$ in (\ref{2-body}),
can be fixed by dimer photoassociation 
experiments such that the trimer photoassociation rate is a prediction of 
this theory.

As an alternative approach to fix $L_2(\Lambda)$, we can study the shift of
the dimer binding energy due to a small change in the static magnetic field.
On one hand,
\be \label{dE1}
\delta E_2 = \bra \psi_B | H_I^{(2b)} | \psi_B \ket =
\frac{L_2}{\Lambda^3}\frac{3}{\pi a} \bra s_0 \ket 
\Lambda^2 \delta B + O(\Lambda).
\ee
On the other hand, in the vicinity of a Feshbach resonance
\be
a(B) \approx a_{bg}\frac{\Delta}{B-B_0},
\ee
where $a_{bg}$ is the background scattering length, $\Delta$ is the resonance
width and $B_0$ its position, therefore
\be
\delta E_2 = \label{dE2}
\frac{dE}{da} \frac{da}{dB} \delta B =
\frac{2\hbar^2}{m a a_{bg} \Delta} \delta B.
\ee
Comparing (\ref{dE1}) and (\ref{dE2}), $L_2(\Lambda)$ can be determined,
\be \label{l2fix}
L_2(\Lambda)= \frac{2\pi}{3} \frac{\hbar^2}{m a_{bg} \Delta}
\frac{1}{\langle s_0 \rangle} \Lambda.
\ee

\begin{figure}
\includegraphics[width=8.6cm]{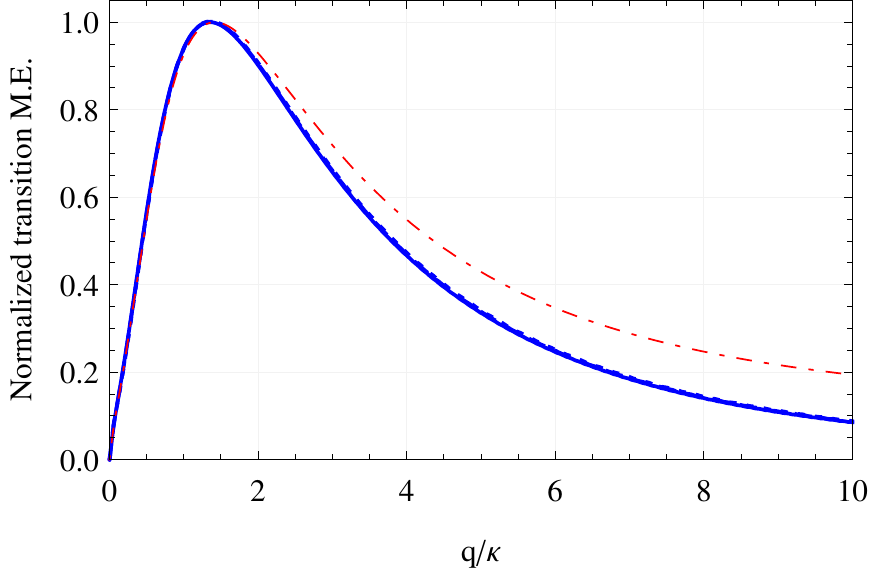}
\caption{\label{DME} (Color online) 
Two-body current: The transition matrix element $I_{2b}^2$, 
as a function of the relative momentum $q/\kappa$,
normalized to 1 for the maximum value at $|a|=\infty$.
For the unitary point the exact result is shown (blue solid line), as well 
as the approximation of (\ref {I2b}) (red dot-dashed line).
Also shown are results for finite scattering length, 
$\kappa a=-100$ (dotted lines) and $\kappa a=-50$ (dashed lines).}
\end{figure}

In Fig. (\ref{DME}) the two-body transition matrix element is shown as a
function of the relative momentum $q/\kappa$.
The validity of the approximation deriving 
(\ref{IJmx}) is also checked by evaluating the integral numerically, starting
from $\rho_0$. It can be seen that while for the low momentum regime this
approximation is excellent, for the high momentum tail it deviates from the 
exact result. 
Numerical results for finite $a$ are also shown in Fig. (\ref{DME}).
For $\kappa a=-100$ and $\kappa a=-50$, we see no significant difference
comparing to the unitary point.

\section{Transition rates}
To evaluate the trimer photoassociation rate (\ref{golden_rule}),
we have to average on the initial states $\bar{\sum_i}$ and to sum over the 
final states $\sum_f$.
The initial three-body state $|i\ket=|qLMl_xl_y\ket$ describes 
three atoms in the continuum with relative momentum $q$ and angular 
momenta $L, M, l_x, l_y$.
The average on these states takes the form
\be \label{sumi}
\bar{\sum_i}=
\frac{1}{\pi} \int_0^\infty dq \sum_{[K]}
\left( R+\frac{d\delta}{dq}\right) P_{[K]}(q) \;, 
\ee
where $P_{[K]}(q)$ is the probability of finding an atomic trio with
quantum numbers $[K]=\{K L M l_x l_y\}$ and momentum $q$.
For system in thermal equilibrium with temperature 
$T$ higher than the condensation temperature,  
$$ P_{[K]} (q) = \frac{1}{\mathcal{Z}}e^{-\frac{\hbar^2 q^2}{2mk_BT}},$$
where $\mathcal{Z}$ is the partition function, $k_B$ is the Boltzmann constant 
and $T$ is the temperature. 

The sum $\sum_f$ contains all possible trimer
quantum numbers and all possible photons weighted by the emission function, 
\be \label{sumf}
\sum_f= \sum_{\bk,\zeta} \sum_{L',M'} ({1+N_{\bk\zeta}}),
\ee
where $N_{\bk\zeta}$ is the number of photons with
momentum $\bk$ and polarization $\zeta$ in the final state.
The stimulating rf radiation is a narrow distribution centered at some 
$\bk_\mathrm{rf}$. Therefore 
$N_{\bk\zeta} \approx N_{\bk_\mathrm{rf}\zeta_\mathrm{rf}}\delta_{\bk, \bk_\mathrm{rf}}
\delta_{\zeta,\zeta_\mathrm{rf}}$,
where $\bk_\mathrm{rf}$ is the momentum of the stimulating rf field
and $\zeta_\mathrm{rf}$ is its polarization.

\subsection{Transition rate in the normal hierarchy case}

Substituting (\ref{t-frozen}), (\ref{sumi}) and (\ref{sumf}) into 
Fermi's golden rule (\ref{golden_rule}), the trimer formation rate in the normal
hierarchy case is given by
\be
r^{1b}_{i\rightarrow f}=
\frac{4 \pi \mu_1^2R}{\epsilon_0 c^6} 
\frac{m}{\hbar^2} 
n_\mathrm{rf}
\ev{s_0} \eta^2 \frac{\omega^5}{q} P(q)
\left ( \frac{I_s^2}{6^2}+\frac{5I_d^2}{15^2} \right ).
\ee
where $n_\mathrm{rf}=N_{\bk_\mathrm{rf},\zeta_\mathrm{rf}}/\Omega$ is the photon
density of the rf field.
The relative momentum $q$ is connected to the photon energy through 
energy conservation $\hbar^2 q^2/2m=-E_3+\hbar \omega$, where $E_3$ is
the trimer binding energy.

\begin{figure}\begin{center}
\includegraphics[width=8.6cm,trim=0cm 0cm 0cm 0mm,clip=true]{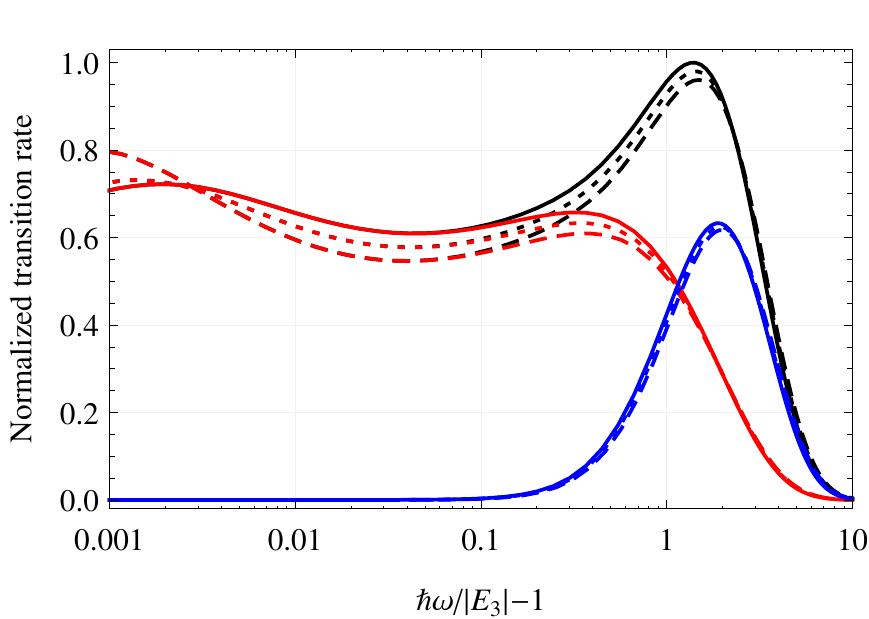}
\includegraphics[width=8.6cm,trim=0cm 0cm 0cm 0mm,clip=true]{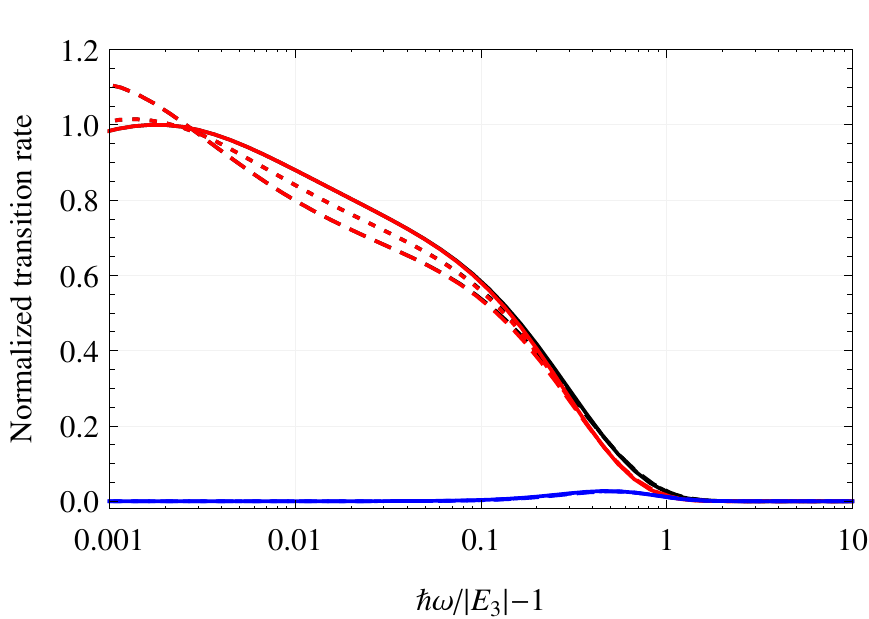}
\caption{\label{fig:rates} (Color online) One-body current: The normalized
trimer photoassociation rate as a function of the rf photon frequency, for 
different temperatures and scattering lengths.
The total (black), the $r^2$ (red, low energy peak) and the quadrupole (blue) 
rates are presented, for temperature of $k_B T=E_3$ (upper panel) 
and $k_B T=0.2E_3$ (lower panel).
Shown are results for the unitary point (solid lines) 
and for finite scattering length, 
$\kappa a=-100$ (dotted lines) and $\kappa a=-50$ (dashed lines).}
\end{center}
\end{figure}

The relative importance of the $s,d$ modes shifts with temperature. 
This point is demonstrated in Fig. \ref{fig:rates}, where the $s,d$ rates
to the trimer photoassociation, at the unitary point $|a|=\infty$ 
and for large but finite scattering length 
are presented for $k_B T=E_3$ and $k_B T=0.2 E_3$. 
The importance of the $d$ mode grows with the ratio $k_B T/E_3$. 
Therefore we can conclude that for small $k_B T/E_3$ values,
the photoassociation is an $s$-wave process while for large
values the two modes have similar importance.
In addition, oscillations in the response can be clearly seen at the low 
frequency regime. These oscillations result form the log-periodic structure of 
the matrix element \cite{BazBar14}. At higher temperature these oscillations are
stronger while at lower temperatures this manifestation of
the Efimov effect is suppressed by the Boltzmann factor.
The amplitude of the oscillations is larger for finite scattering length
than at the unitary point.

\subsection{Transition rate in the strong hierarchy case}

For the strong hierarchy case, the trimer formation rate is given by
\be \label{StrongTrimer}
r^{2b,t}_{i\rightarrow f}=
\frac{L_2^2}{\Lambda^6} \frac{4 m R}{\hbar^2 c^2 \epsilon_0} 
n_\mathrm{rf}
\ev{s_0}\eta^2 \frac{\omega}{q} P_t(q) I_{2b,t}^2.
\ee
Here we have added the subscript ``$t$'' to $P(q), I_{2b}$ to
stress these are trimer related quantities.
The corresponding dimer formation rate for this case is given by
\be \label{StrongDimer}
r^{2b,d}_{i\rightarrow f}=
\frac{L_2^2}{\Lambda^6} \frac{2 m R}{\hbar^2 c^2 \epsilon_0} 
n_\mathrm{rf}
\ev{s_0}\eta^2 \frac{\omega}{q} P_d(q) I_{2b,d}^2,
\ee
where $P_d(q)$ is the probability for a pair of particles to be in
$L=0$ state with momentum $q$. 
Comparing these two results and using Eqs. (\ref{i2bd}),(\ref{l2fix})
we finally obtain
\be \label{rratios}
\frac{r^{2b,t}_{i\rightarrow f}}{r^{2b,d}_{i\rightarrow f}}
   =2\frac{P_t(q)}{P_d(q)}\frac{I_{2b,t}^2}{I_{2b,d}^2}
\ee
and
\be\label{r2fixed}
{r^{2b,d}_{i\rightarrow f}}=
\left(\frac{\hbar^2}{m a_{bg} \Delta}
 \right)^2
\frac{2 m }{\hbar^2 c^2 \epsilon_0} 
     n_\mathrm{rf}
     \eta^2 \frac{\omega}{q} P_d(q) \kappa
        \sin^2 \delta(q)
\ee
One should note that these results are independent of
the cutoff parameter $\Lambda$. The dependence of the rates
on the normalization radius $R$ enters
through the probability $P_{d/t}(q)$.

\begin{figure}\begin{center}
\includegraphics[width=8.6cm,trim=0cm 0cm 0cm 0mm,clip=true]{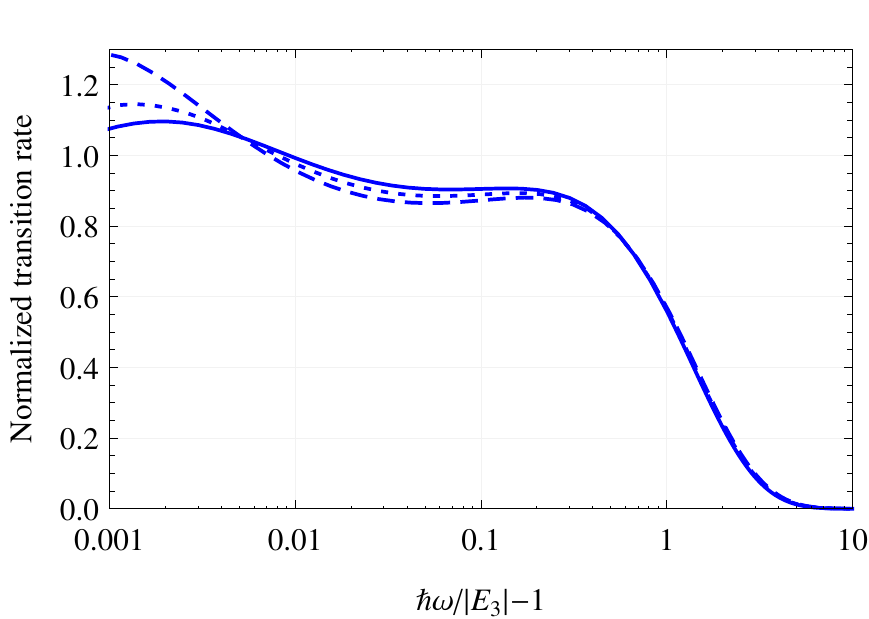}
\caption{\label{RateDelta} (Color online) Two-body current: The normalized 
trimer photoassociation rate as a function of the rf photon frequency.
Shown are the results for $k_B T=E_3$ at the unitary point 
(solid lines) and at finite scattering lengths, $\kappa a=-100$ (dotted line)
and $\kappa a=-50$ (dashed line).}
\end{center}
\end{figure}

The trimer transition rate for the strong hierarchy case is shown for
$k_B T=E_3$ in Fig. \ref{RateDelta}. Here as in the normal hierarchy case, 
log periodic oscillations are visible. For finite scattering length they
are amplified. Since there is only a single channel here, the low temperature 
behavior is similar to the high temperature, and therefore not shown.

\section{Experimental results}

In this section we apply our theory to the photoassociation of ultracold
$^7$Li atoms, and fit it to the experimental results of the Bar-Ilan group
\cite{MacShoGro12}.

In this experiment, a gas of $^7$Li atoms on the
$|F=1,m_F=0\ket$ state was cooled down to $T\approx 1.5\mu K$,
with mean density of about $10^{12} \mathrm{cm}^{-3}$.
A static magnetic field was applied in the vicinity of a Feshbach resonance
at $\sim$894 G.
Then an rf field was turned on for $\sim1$ msec. The magnetic component
of the rf field was parallel to the static magnetic field, therefore $\eta=1$. 
The number of atoms remains in the trap was measured as function of the 
rf frequency.

The scattering length in this experiment is positive, therefore both dimers and
trimers can be associated by the rf field.
The experimental signal for photoassociation is particle loss from the trap.
In both cases three particles are lost for each associated molecule.
Internal collisions break the trimer into a deeply bound
dimer and an atom, both carry high kinetic energy and therefore 
escape the trap.
The dimer collides with a third atom and
decays to a deeper bound state, and in the process enough energy 
is released such that both
atom and dimer escape the trap.

To connect the photoassociation rate to the measured quantity  
we integrate the population evolution equation
\be
  \dot N = -3 r^{2b,d}_{i\rightarrow f} \frac{N^2}{2}
           -3 r^{2b,t}_{i\rightarrow f} \frac{N^3}{3!} \;,
\ee
where the dimer (trimer) photoassociation rate is multiplied by the numbers of
pairs (trios) is the system.
Since this experiment is clearly in the strong hierarchy regime, we use
Eqs. (\ref{StrongTrimer}) and (\ref{StrongDimer}) for the transition rates.

To fix the value of $L_2$ we use Eq. (\ref{l2fix}), where the relevant Feshbach
resonance parameters are
$a_{bg} \approx -18.24 a_0$ and $\Delta \approx -237.8$G
\cite{GroShoMac11}. Doing so, see Eqs. (\ref{rratios}),(\ref{r2fixed}), 
the transition rate is independent
of $\Lambda$ and also of the effective charge $\bra s_0 \ket$.

The photon density is connected to the amplitude of the oscillating magnetic
field $B_\mathrm{rf}$
through $n_\mathrm{rf}= B_\mathrm{rf}^2/2\mu_0\hbar\omega$, where $\mu_0$ is the
vacuum permeability. 

Due to strong rf field there is power broadening 
which blurs the fine structure of the transition rate. To account for
this effect we convolute the transition rates with a Gaussian,
\be
r_{i\rightarrow f}(\omega)=\frac{1}{\sqrt {2 \pi } \sigma}
\int d\omega' r_{i\rightarrow f}(\omega') e^{-\frac{(\omega-\omega')^2}{2 \sigma^2}},
\ee
where the Gaussian width $\sigma$ was fitted to the experimental results. 

The fitting parameters are shown in Table \ref{FitParm}.

\begin{table}
\caption{\label{FitParm} The fitting parameters for the $^7$Li
  photoassociation experimental results.
  Here $n$ is the number density and $\sigma$ is the broadening width.}
\begin{indented}
\item[]\begin{tabular}{@{}cccccc}
\br
Case & $a (a_0)$&$n (10^{12}\mathrm{cm}^{-3})$&$B_\mathrm{rf}$ (G)&$E_3/h$ (MHz)& $\sigma$ (kHz) \\
\mr
a & 806 & 1.3 & 0.6  & 0.93 & 25 \\
b & 629 & 1.0 & 0.52 & 1.45 & 20 \\
\br
\end{tabular}
\end{indented}
\end{table}

\begin{figure}\begin{center}
 \includegraphics[width=8.6cm,trim=0cm 0cm 0cm 0mm,clip=true]{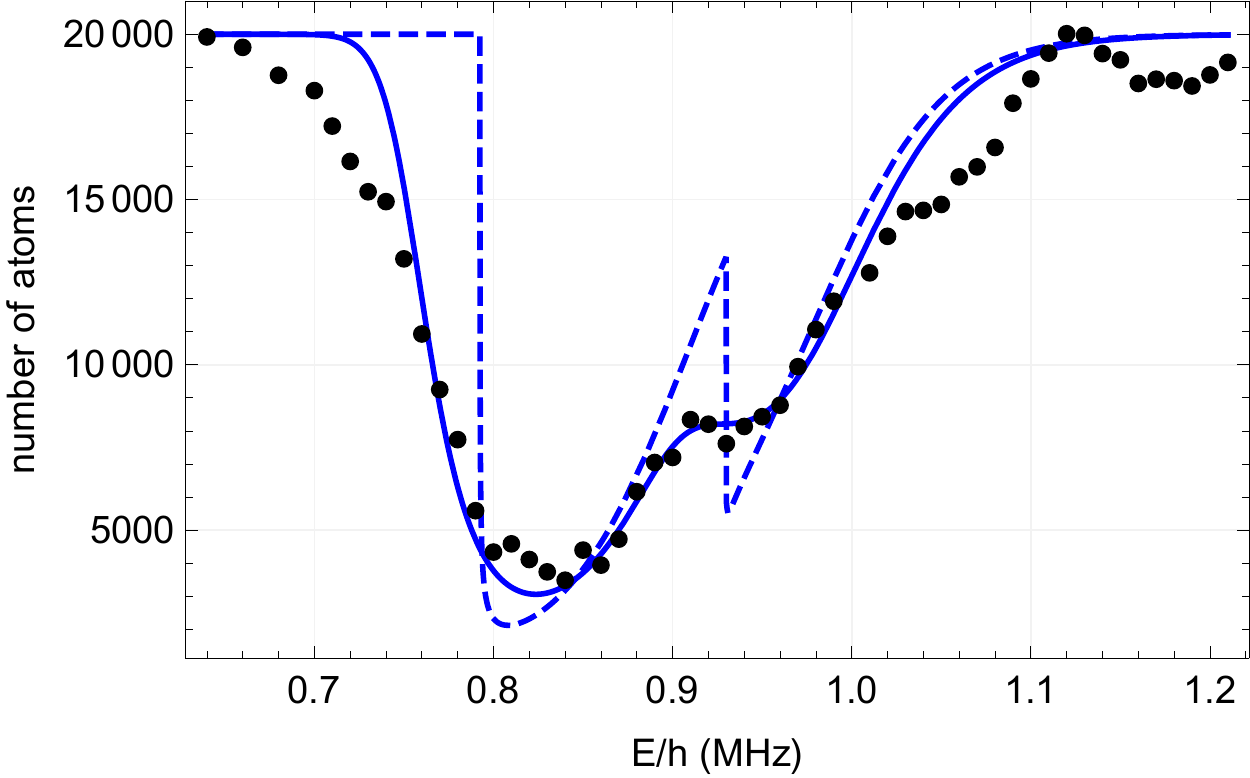}
 \includegraphics[width=8.6cm,trim=0cm 0cm 0cm 0mm,clip=true]{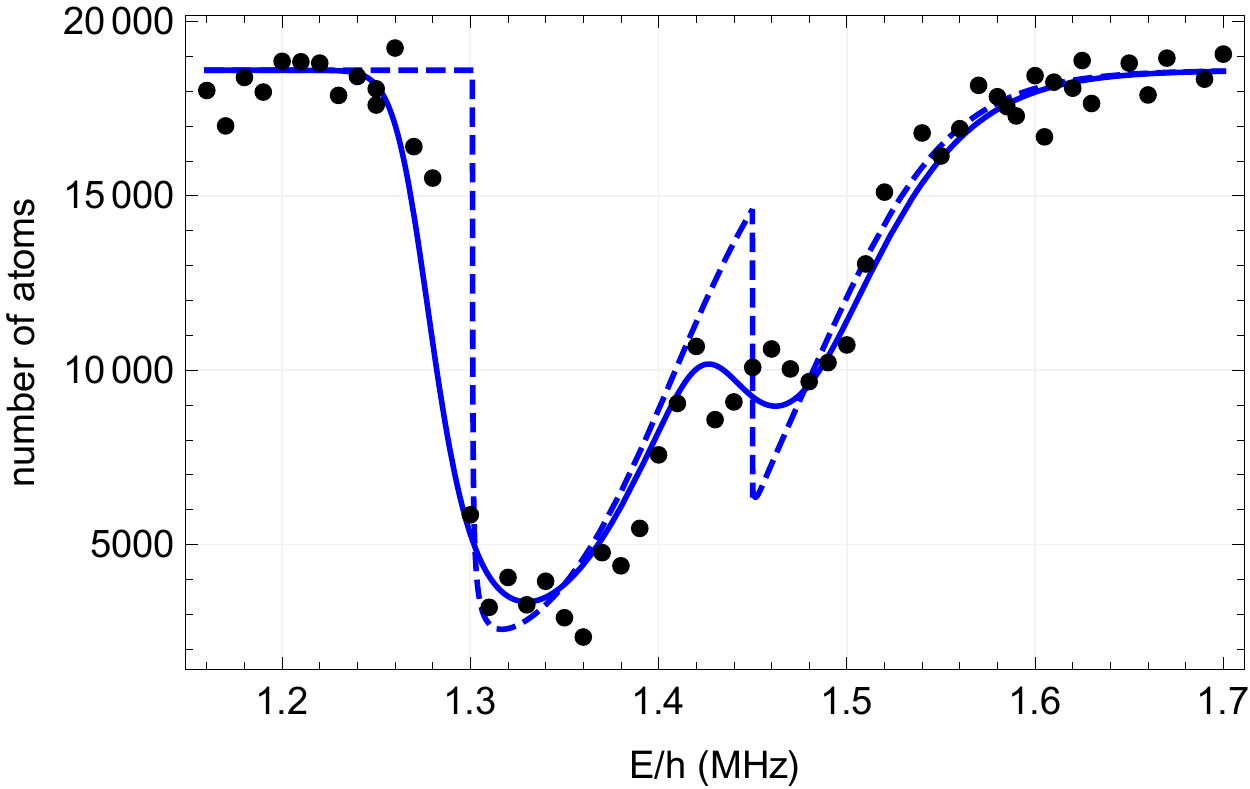}
\caption{\label{ExpFit} (Color online)
  The number of $^7$Li atoms remaining in the trap after photoassociation
  experiment, as function of the rf frequency.
  Shown are the experimental results from Ref. \cite{MacShoGro12},
  for $a/a_0 \approx 810$ (upper plot) and $a/a_0 \approx 627$ (lower plot),
  as well as our fitting results.
  To distinguish the power broadening effect the results without broadening
  are also shown in dotted curve.
}
\end{center}
\end{figure}

Our theory for photoassociation of $^7$Li ultracold atoms, as well as 
the experimental results of Ref. \cite{MacShoGro12} are shown in
Fig. \ref{ExpFit}.
To distinguish the power broadening effect the results without broadening
are also shown in dotted curve.
Our theory agrees well with the experimental results, and the fitting
parameters are within the experimental uncertainty. However,
the experimental resolution limits the ability to identify fine
details of the association process.

\section{Conclusion}
To conclude, we have applied the multipole expansion to study
universal trimer photoassociation.
One-body and two-body current were explored.
For normal hierarchy case, where the one-body current dominates, the 
two leading $s-$ and $d-$ modes are studied and their relative 
contribution is shown to vary with temperature.
In ultracold atoms, and other strong hierarchy systems, the relevant 
operator is found to be the two-body current. The log-periodic oscillations
which modulate the cross section in both scenarios is found to be 
amplified for finite scattering length compared to the unitary point.
We apply out theory to $^7$Li ultracold atoms photoassociation and
show nice fit with the available experimental results.
\ack
This work was supported by the Pazi fund.
We thank the Institute for Nuclear Theory at the University of Washington 
for its hospitality and the US Department of Energy for partial support during 
part of the work presented here.
We would like to thank L. Khaykovich for useful discussions and for providing
us his experimental results.

\appendix

\section{Raynal-Revai coefficients} \label{RayRev}

Our calculation of the projector operator (\ref{Rot}), provides 
an explicit formula for the $l_x=0$ or $l_y=0$ Raynal-Revai coefficients 
$\bra l_x l_y | l'_x l'_y \ket_{KL}$ \cite{RayRev70}, defined by
\be
\mathcal Y^{KLM}_{l_xl_y}(\Omega_i)=
\sum_{l'_xl'_y} \bra l_x l_y | l'_x l'_y \ket_{KL} 
\mathcal Y^{KLM}_{l'_xl'_y}(\Omega_j).
\ee
Using the orthogonality of the hyperspherical harmonics we get, 
\be
\bra l_x l_y | l'_x l'_y \ket_{KL}=
\frac{R_{ij}^{(l_xl_y)(l_x'l_y')}[\varphi^K_{l'_xl'_y}(\alpha_j)](\alpha_i)}
{\varphi^K_{l_xl_y}(\alpha_i)}.
\ee
Putting $l_x=0, l_y=L$ and $\alpha_i=0$,
\be\eqalign{
\bra 0 L | l'_x l'_y \ket_{KL}=
(\mp)^{l'_x}\sqrt{(2l'_x+1)(2l'_y+1)}\times 
\cr \threej {l_x'}{l_y'}{L}{0}{0}{0}
\frac{\varphi^K_{l'_xl'_y}(\pi/3)}{\varphi^K_{0 L}(0)}.}
\ee
Similar calculation for $l_x=L, l_y=0$ and $\alpha_i=\pi/2$ gives
\be\eqalign{
\bra L 0 | l'_x l'_y \ket_{KL}=
(\pm)^{l'_y}\sqrt{(2l'_x+1)(2l'_y+1)}\times 
\cr \threej {l_x'}{l_y'}{L}{0}{0}{0}
\frac{\varphi^K_{l'_xl'_y}(\pi/6)}{\varphi^K_{L 0}(\pi/2)}.}
\ee

\section{The hyperangular integrals} \label{Rotation}

The wave function is written as sum of Faddeev components, each
in different coordinate set,
\be
\Phi_\nu^L(\Omega)=\frac{1}{\sqrt {4\pi}}\sum_{i=1}^3 Y_{L0}(\hat y_i)
 P_\nu^L(\alpha_i)/\sin 2\alpha_i.
\ee 
Our aim is to evaluate the hyperangular integral,
\be\label{Iha2}
I_\Omega=\int d\Omega_1 \Phi^{0*}_\nu(\Omega) 
\left(\sum_j Y_{20}(\hat y_j)\cos^2\alpha_j\right)
\Phi^2_{\nu'}(\Omega) 
\ee
Due to symmetry, one of the sums, say over $j$,
can be dropped giving a factor of 3. The integral (\ref{Iha2}) can be
further simplified
utilizing the body-fixed coordinate system  \cite{BarEfrLei04}.  
Consider the two body-fixed vectors $\bff x_{10}$ and $\bff y_{10}$ such that
$\hat x_{10} \cdot \hat y_{10}=\gamma_i$. The transformation from the body-fixed to
system to the laboratory system is given through the relation
\be
Y_{l m}(\hat y_i)=\sum_{\mu}D_{\mu m}^{l}(\omega)Y_{l \mu}(\hat y_{i0}),
\ee
where $D_{\mu m}^{l}(\omega)$ is the Wigner D-matrix, and
the rotation $\omega$ is the set of 3 Euler angles.
Applying this transformation, for fixed $\alpha_i$,
\be\eqalign{
\frac{1}{4\pi}&\int d\hat x_i\int d\hat y_i Y_{l m}^*(\hat y_i) Y_{l' m'}(\hat y_j)=
\\ \nonumber 
\int d\gamma_i& \sum_{\mu,\mu'} Y_{l \mu}^*(\hat y_{i0}) Y_{l' \mu'}(\hat y_{j0})
\int d\omega D_{\mu m}^{l*}(\omega) D_{\mu' m'}^{l'}(\omega)}.
\ee
Now one can utilize the orthogonality of the Wigner D-matrix,
\be
\int d\omega D_{\mu m}^{l*}(\omega) D_{\mu' m'}^{l'}(\omega)=
\frac{8\pi^2}{2l+1}\delta_{l,l'}\delta_{\mu,\mu'}\delta_{m,m'}
\ee
and the addition theorem 
\be
\sum_{\mu} Y_{l \mu}^*(\hat y_{i0}) Y_{l \mu}(\hat y_{j0})=
\frac{2l+1}{4\pi}P_l(\gamma_{ij})
\ee  
where $P_l(\gamma_{ij})$ is the Legendre polynomials and 
$\gamma_{ij}=\hat y_{i0}\cdot \hat y_{j0}$.
Choosing 
$\bff x_{i0}=\rho \sin(\alpha_i)\hat z$ and 
$\bff  y_{i0}=\rho \cos(\alpha_i)(\sqrt{1-\gamma_i^2}\hat x+\gamma_i\hat z)$,

\be
\gamma_{ij}(\alpha_i,\gamma_i)=
-\frac{\cos \alpha_i \pm \sqrt 3 \gamma_i \cos \alpha_i}
{\sqrt{1+2\sin^2\alpha_i\pm \sqrt 3 \gamma_i \sin 2\alpha_i }}.
\ee
Finally,
\be\eqalign{
I_\Omega=\frac{3}{4\sqrt \pi}
\int_{-1}^1 d\gamma_1 \int_0^{\pi/2} d\alpha_1 
\sin^2 \alpha_1 \cos^4 \alpha_1 \times \\
\sum_{i,k}P_2(\gamma_{1k}) \frac{P_\nu^{0*}(\alpha_i)}{\sin 2\alpha_i}
                        \frac{P_{\nu'}^2(\alpha_k)}{\sin 2\alpha_k},}
\ee
where $\alpha_i=\alpha_i(\alpha_1,\gamma_1)$ is given by (\ref{alpha}).


\end{document}